\documentstyle[epsf]{elsart}

\begin {document}
\begin{frontmatter}

\title {A new analysis method 
for very high definition Imaging Atmo\-spheric Cherenkov Telescopes
as applied to the CAT telescope }

\author[cdf]{S. Le Bohec}
\author[x]{B. Degrange}
\author[cdf]{M. Punch}
\author[paris]{A. Barrau}
\author[cesr]{R. Bazer-Bachi}
\author[perpi]{H. Cabot\thanksref{dead}}
\author[x]{L.M. Chounet}
\author[perpi]{G. Debiais}
\author[cesr]{J.P. Dezalay}
\author[paris]{A. Djannati-Ata\"{\i}}
\author[cenbg]{D. Dumora}
\author[cdf]{P. Espigat}
\author[perpi]{B. Fabre}
\author[x]{P. Fleury}
\author[x]{G. Fontaine}
\author[paris]{R. George}
\author[cdf]{C. Ghesqui\`ere}
\author[sap]{P. Goret}
\author[sap]{C. Gouiffes}
\author[sap,paris7]{I.A. Grenier}
\author[x]{L. Iacoucci}
\author[cesr]{I. Malet}
\author[perpi]{C. Meynadier}
\author[cdf,czech]{F. Munz}
\author[purd]{T.A. Palfrey}
\author[x]{E. Par\'e}
\author[paris]{Y. Pons}
\author[cenbg]{J. Qu\'ebert}
\author[cenbg]{K. Ragan}
\author[sap,paris7]{C. Renault}
\author[paris]{M. Rivoal}
\author[czech]{L. Rob}
\author[czopt]{P. Schovanek}
\author[cenbg]{D. Smith}
\author[paris]{J.P. Tavernet}
\author[x]{J. Vrana\thanksref{dead}}

\address [cenbg]{Centre d'Etudes Nucl\'eaire de
Bordeaux-Gradignan, France\thanksref{in2p3}}
\address [cesr]{Centre d'Etudes Spatiales des
Rayonnements, Toulouse, France\thanksref{insu}}
\address [cdf]{
Laboratoire de Physique Corpuscul\-aire et Cosmologie,
Coll\`ege de France, Paris, France\thanksref{in2p3}}
\address [x]{
Laboratoire de Physique Nucl\'eaire de Haute Energie, 
Ecole Polytechnique, Palaiseau, 
France\thanksref{in2p3}}
\address [paris]{LPNHE,
Lab.oratoire de Physique Nucl\'eaire de Haute Energie,
Universit\'es de Paris VI/VII, France\thanksref{in2p3}}
\address [sap]{Service d'Astrophysique, Centre d'Etudes
de Saclay, France\thanksref{cea}}
\address [perpi]{Groupe de Physique Fondamentale,
Universit\'e de Perpignan, France\thanksref{in2p3}}
\address [czech]{Nuclear Center, Charles University, Prague, Czech
Republic}
\address [paris7]{Universit\'e Paris VII, France}
\address [purd]{Department 
of Physics, Purdue University, Lafayette, IN
47907, U.S.A.}
\address [czopt]{Joint Laboratory 
of Optics Ac. Sci. and Palacky
University, Olomouc, Czech Republic}

\thanks [in2p3]{IN2P3/CNRS}
\thanks [insu]{INSU/CNRS}
\thanks [cea]{DAPNIA/CEA}
\thanks [dead]{Deceased}

\begin{abstract}
A new method of shower-image analysis is presented which appears very
powerful as applied to those Cherenkov Imaging Telescopes with very
high definition imaging capability. It provides hadron rejection on the
basis of a single cut on the image shape, and simultaneously determines
the energy of  the electromagnetic shower and the position of the
shower axis with respect to the  detector. The source location is also
reconstructed for each individual $\gamma$-ray shower, even with one
single telescope, so for a point source the hadron rejection can be
further improved. As an example, this new method is  applied to data
from the C{\small AT} (Cherenkov Array at Th\'emis) imaging  telescope,
which has been operational since Autumn, 1996.
\end{abstract}
\begin{keyword}
Gamma-Ray Astronomy, Atmospheric Cherenkov detector
\end{keyword}
\end{frontmatter}

\section { Introduction }

Most of the recent progress in ground-based $\gamma$-ray Astronomy has
been obtained from  Atmospheric Cherenkov Telescopes ({\small ACT})
using the imaging technique, which permits efficient rejection of the
large background of proton or nucleus-induced showers. The utility of
high-definition imaging has been demonstrated by the results obtained
with the Whipple (109 pixels) \cite{whitel}, C{\small AN\-GAROO} (220
pixels) \cite{cantel}, and  H{\small EGRA} (271 pixels) \cite{hegratel}
{\small ACT}'s, leading to the discovery of several  firmly established
very-high-energy sources;  e.g., the Crab nebula \cite{whicrab}, the
pulsar {\small PSR} 1706-44 \cite{canpsr}, and the two relatively
nearby Active Galactic Nuclei, Markarian 421 \cite{whi421} and 
Markarian 501 \cite{whi501}. Using a camera with a large number of
small pixels allows a more profound image analysis than the usual
procedure based on the first and second moments of the light
distribution \cite{scuts}, with improved background rejection and
energy resolution. The new image analysis  described here takes
advantage of the full imaging information, including the detailed 
longitudinal development of the shower (with a significant asymmetry
between the top and the bottom of the shower) as well as its lateral
extension.  This method is based on an analytical model giving, for a
genuine $\gamma$-ray shower, the average distribution of Cherenkov
light in the focal plane as a function of the following parameters: 
$\gamma$-ray energy $E_{\gamma}$,  distance $D$ between the shower axis
and the telescope (or impact parameter), the source position (defined
by the vector $\vec{\xi}$ in the focal plane),  and the azimuthal
position $\phi$ of the shower image about the source in the focal
plane. This model is used to define a $\chi^{2}$-like function of
$E_{\gamma}$,  $D$, $\vec{\xi}$, and $\phi$.  For each observed shower,
this function is minimized with respect to the parameters: the
minimized $\chi^{2}$ provides gamma-hadron discrimination;  the
$\gamma$-ray  energy determination automatically takes account of the
impact parameter which is obtained from the same fit.  With this
method, the origin of a genuine $\gamma$-ray shower can be
reconstructed  even with one single telescope.  This result stems from
the fact that electromagnetic showers whose energies are known have a
well-defined longitudinal development with rather small fluctuations. 
In the focal plane of an imaging {\small ACT}, this gives a
well-defined longitudinal profile for the image, which depends on the
impact parameter and angular origin of the shower.  Therefore, the
simultaneous measurement of $E_{\gamma}$ and $D$ in the preceding fit
also yields the source position $\vec{\xi}$. For the case of a
well-localized point source, the energy resolution can be improved by
fixing the source position during the fit since there are fewer
parameters in the fit.

In this article, the preceding method is illustrated as applied to the 
C{\small AT} imaging telescope \cite{cat97}, operating since Autumn
1996 in the French Pyrenees.  Its very-high-definition imaging camera
comprises a central region consisting of 546 phototubes in a hexagonal
matrix spaced by 0.13$^\circ$; this is surrounded by 54 tubes in two
``guard rings''. The angular diameter of the full field of view is
$4.8^\circ$  ($3.0^\circ$ for the small pixels). The guard rings were
installed in June 1997,  so for the data and simulations used here,
they are not included.  A more complete description of the C{\small
AT}  imaging telescope can be found in \cite{cat97}. The detector
response has been simulated in detail, thus allowing the optimization
of the new image analysis on the basis of Monte-Carlo-generated
$\gamma$-showers against the background of real off-source data.  

Section 2 is devoted to the full simulation of the  instrument,
allowing the calculation of the variation of the equivalent detection
area as a function of $E_{\gamma}$ for different zenith angles, thus
defining the corresponding energy thresholds. In Section 3, the
semi-analytical model giving the average Cherenkov light distribution
in the focal plane for a $\gamma$-ray shower is described in detail and
shown to be consistent with results from complete simulations of
electromagnetic cascades.  In Section 4, the performance of the
analysis method is presented for different values of the zenith angle;
the discussion is focused on hadron rejection by a single cut on the
image shape (and for a known point source, an additional cut on
direction),  on the accuracy of the source location, and finally on
energy resolution.

\section {Simulation of the detector response}

The development of electromagnetic or hadronic air showers in the
atmosphere is simulated by a modified version of the program described
in \cite{whimc}.  For each event, the Cherenkov photons falling onto
the mirror elements are followed individually according to their
arrival times, initial directions, and wavelengths; reflection from the
Davies-Cotton mirror is calculated, taking into account imperfections
of the mirrors' shapes and orientations. Typical aberration images from
a point source are smaller than or comparable to the pixel size. A
wavelength-dependent fraction of  photons is kept, reproducing the
effect of mirror reflectivity and Winston  cones' efficiency
\cite{cones}  as well as phototubes' quantum efficiency.  Photons from
the sky-noise are also simulated, taking into account the effect of
{\small AC} coupling on the read-out electronics.  The contribution of
typical star fields may be superposed. The time-structure and amplitude
fluctuation of the pulses generated by single $\gamma$e's have been
parametrized on the basis of tests with the C{\small AT} electronics.
The signal from a given  pixel is obtained  from the pile-up of the
contributions from individual $\gamma$e's according to their respective
arrival times, taking account of the amplitude fluctuation of $\sim 0.4
\gamma$e.  The comparator threshold is set at a level corresponding to
three times the height of the average  single-$\gamma$e signal. When a
trigger occurs, the phototube signal (corrected for the effect of
propagation along the $28\:{\mathrm {m}}$ cable) is integrated over the
$12\:{\mathrm {ns}}$ time window provided by the fast gate, thus
yielding the simulated {\small ADC} output. These {\small ADC} outputs
are then used as input to the analysis procedure described in the
following sections.

\begin{figure}
$$
\epsfxsize=6.2 cm
(a)\epsffile[90 210 510 630]{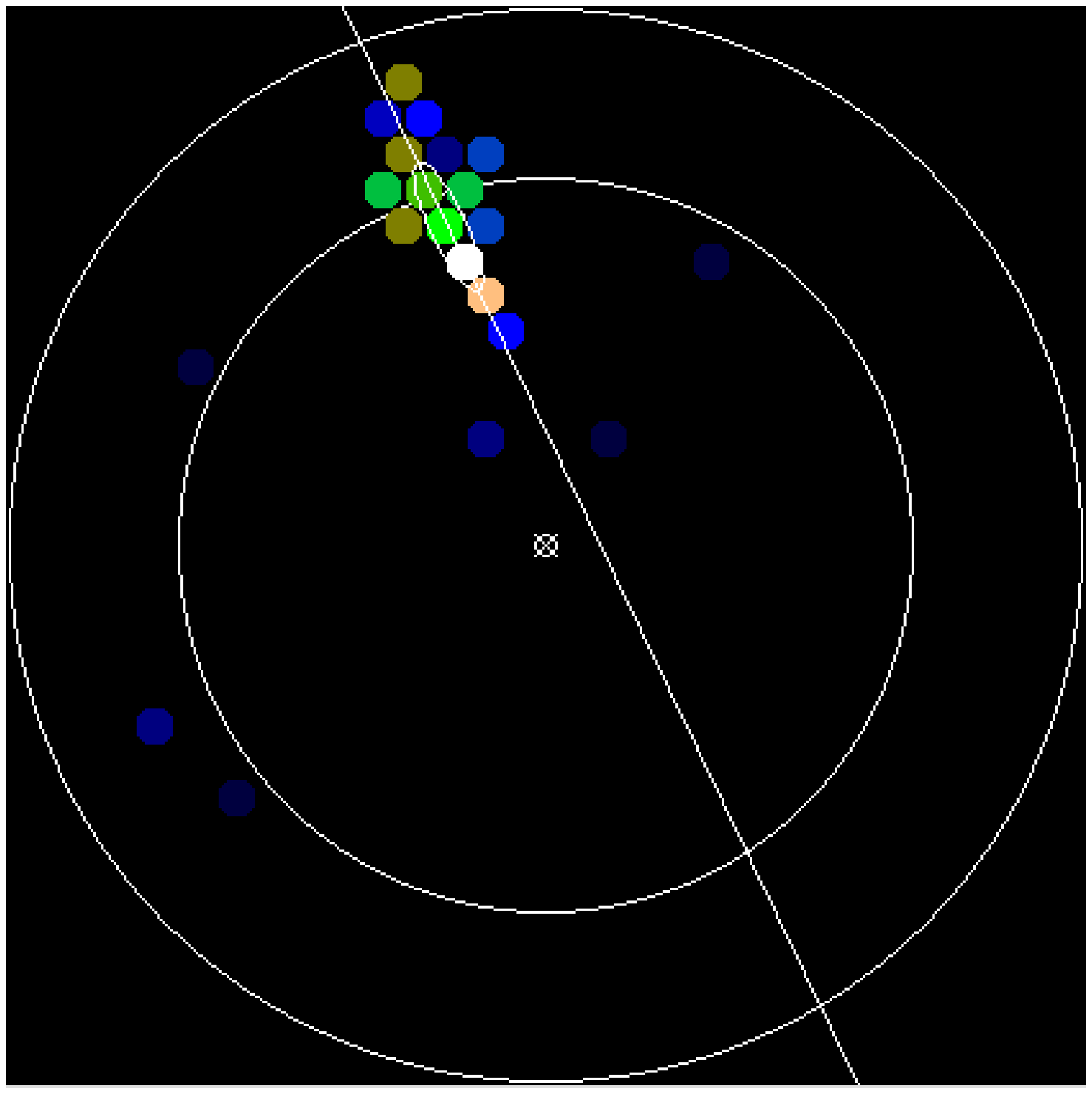}
\;
\epsfxsize=6.2 cm
(b)\epsffile[90 210 510 630]{hadron35.ps}
$$
$$
\epsfxsize=6.2 cm
(c)\epsffile[90 210 510 630]{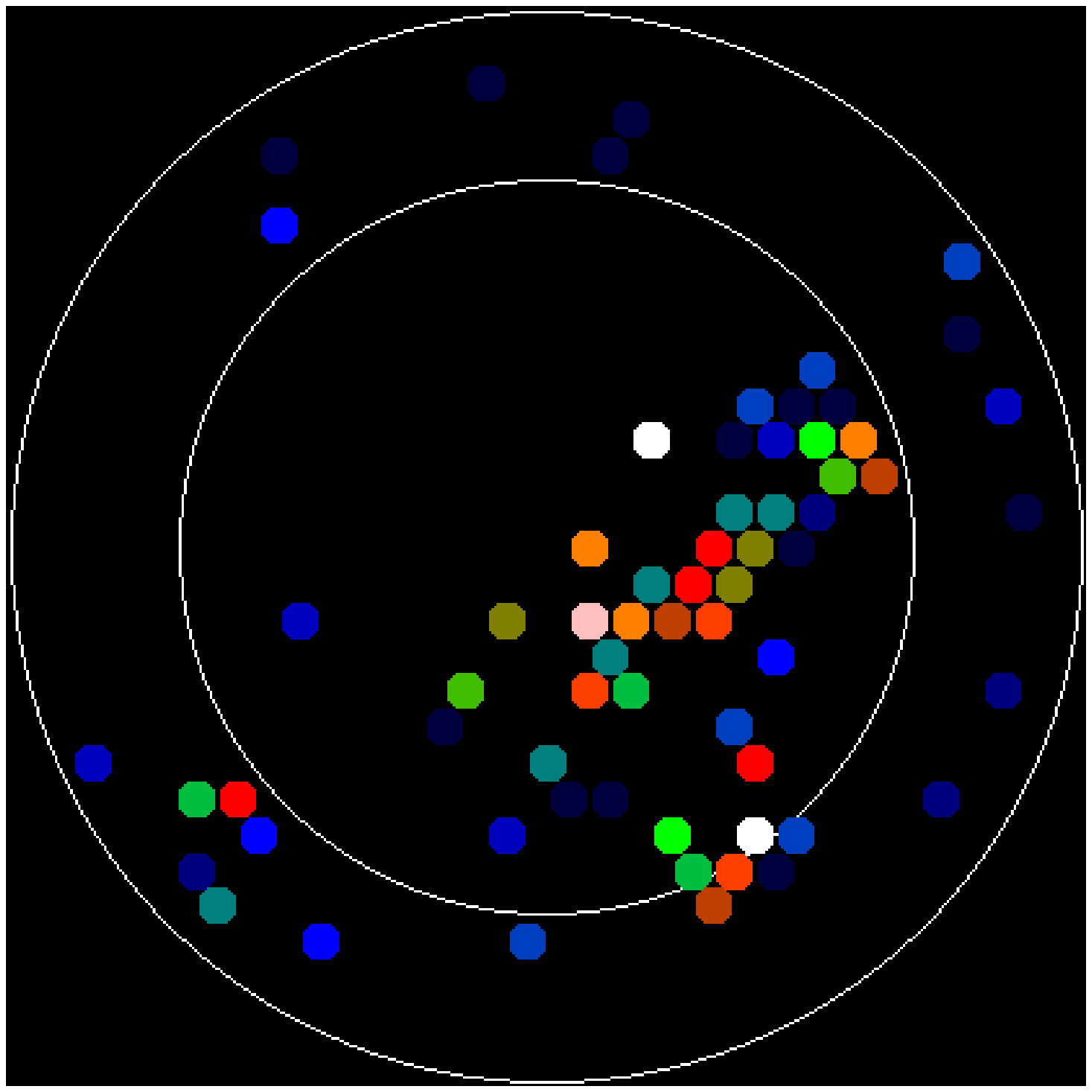}
\;
\epsfxsize=6.2 cm
(d)\epsffile[90 210 510 630]{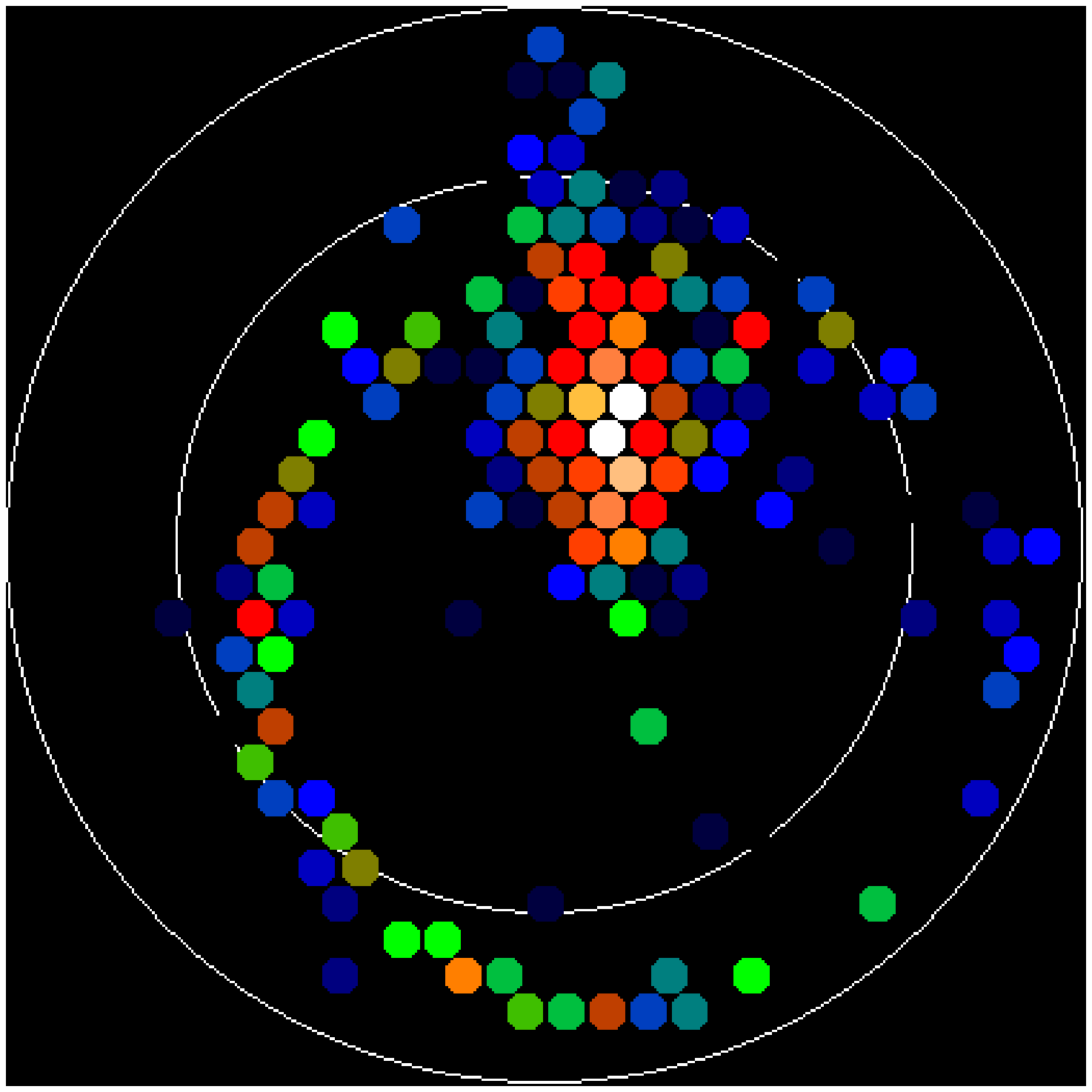}
$$
\caption{ Real shower images in the C{\small AT} imaging telescope: a)
$\gamma$-ray shower candidate (estimated energy $620\:{\mathrm GeV}$); 
b) and c) hadronic images; d) hadronic image with a characteristic muon 
arc. The concentric circles represent the trigger region and the 
small-pixel region, respectively.}
\label{fig:images}
\end{figure}

The preceding simulation of instrumental effects allows the calculation
of the equivalent detection area for $\gamma$-rays as a function of
energy for a given zenith angle $Z$. This is shown in
Fig.~\ref{fig:accept}.a for  $Z$s of $0^{\circ}$, $30^{\circ}$, 
$45^{\circ}$, and $60^{\circ}$. Fig.~\ref{fig:accept}.b shows the
corresponding event rates of $\gamma$-rays for a reference source with
the intensity and spectrum of the Crab nebula, as measured by the
C{\small AT} telescope \cite{catcrab}:
\begin{figure}
\epsfxsize=14.5 cm
\leavevmode
\centering
\epsffile[0 25 580 520]{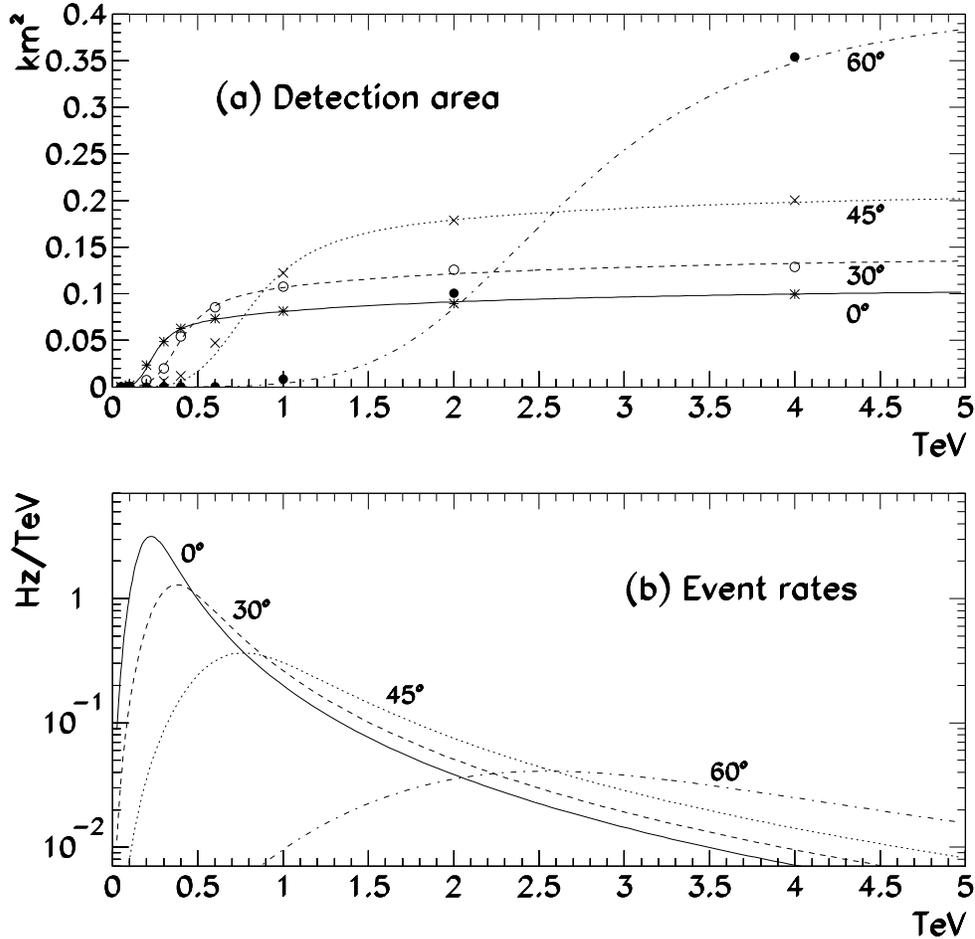}
\caption{ 
a) Equivalent detection area for $\gamma$-rays as a function  of energy
for various values of zenith angle. The points show the values at which
the acceptance was evaluated from  Monte Carlo simulations, the curves
are a fit to these points.
b) Corresponding event rates per energy interval for $\gamma$-rays
from a Crab-like source with a spectrum as given in Equation (1).}
\label{fig:accept}
\end{figure}
\begin{equation}
\frac{d\phi}{dE} = 2.46 \times \left( \frac{E}{\mathrm{TeV}} \right) 
^{-2.55} \times 10^{-11} \; \; \; {\mathrm cm^{-2} s^{-1} TeV^{-1}}
\end{equation}
The energy corresponding to the maximum event rate (``mode energy'')
will be considered as the nominal threshold of the telescope; it varies
from 250 to $350\:{\mathrm {GeV}}$ between zenith and  $Z = 30^{\circ}$
and increases to  $\approx 700\:{\mathrm {GeV}}$ for $Z = 45^{\circ}$. 
Examples of real images from a $\gamma$-ray candidate shower and from
hadron showers  are shown in Fig.~\ref{fig:images}. 
Whereas $\gamma$-rays produce thin elongated images
(Fig.~\ref{fig:images}.a), patterns  from hadron showers are often more
chaotic or patchy due to the superposition of several electromagnetic
components due to $\pi^{0}$'s (Fig.~\ref{fig:images}.b and
\ref{fig:images}.c) and to Cherenkov rings or arcs generated by muons
falling onto the mirror or close ($< 40\:{\mathrm {m}}$) to the
telescope (Fig.~\ref{fig:images}.d). For hadron showers with energies
lower than $200\:{\mathrm {GeV}}$, often the only component which is
seen is the Cherenkov light from a single muon; such muons are a source
of background in the case where the image is reduced to a short arc,
which mimics a low-energy $\gamma$-ray shower \cite{mu93}. It should be
noted, however, that the  C{\small AT} imaging telescope is surrounded
by the seven detectors of the ASGAT timing array \cite{asgat} (whose
collectors are $7\:{\mathrm {m}}$ diameter each) which should permit
the rejection of most of these background events when it resumes
operation after upgrade, in 1998.

\section {Modelling of $\gamma$-shower 
images and image analysis method}

For a given value of energy $E_{\gamma}$ and impact parameter, $D$, the
Cherenkov image of an  electromagnetic shower fluctuates about the mean
image, as a result both of intrinsic fluctuations of the shower and of
statistical effects in the Cherenkov light collection, as well as
instrumental fluctuations. An analysis based on the comparison between
individual event images and the theoretical mean image as a function of
impact parameter and energy allows full advantage to be taken of the
high definition of the  C{\small AT} imaging telescope camera. Such an
analysis requires an analytical model for $\gamma$-shower Cherenkov
images.

In order to construct this model, we have used the results from a paper
by Hillas \cite{hillas} in which the mean development of
electromagnetic showers is described and parametrized on the basis of
Monte-Carlo simulations. The number of charged particles ($e^{\pm}$) at
a given atmospheric depth is given by the Greisen formula. Their energy
spectrum and the angular distribution of their momenta with  respect to
the shower axis are given. Their spatial distribution around the shower
axis is not explicitly stated, but parametrizations of their mean
distance from this axis and of their spreads in the radial and
azimuthal directions are given as a function of their energy and
momentum angle. These parametrizations have been compared with our
Monte-Carlo simulations, and were found to be in very good agreement,
except for the mean value of the angular distribution which we have
modified slightly (see Appendix 1). This description of the mean shower
development permits the mean Cherenkov image to be deduced, with the
additional use of the atmospheric density  profile as in \cite{whimc},
optical absorption as in \cite{xavier},  Cherenkov emission properties,
and some of the  detector characteristics such as its light-collecting
area, photo-tube quantum efficiency as a function of wavelength, and
site altitude. The angular distance from the source to the zenith has
also to be taken  into account.

\begin{figure}
\epsfxsize=14.5 cm
\leavevmode
\centering
\epsffile[28 160 590 695]{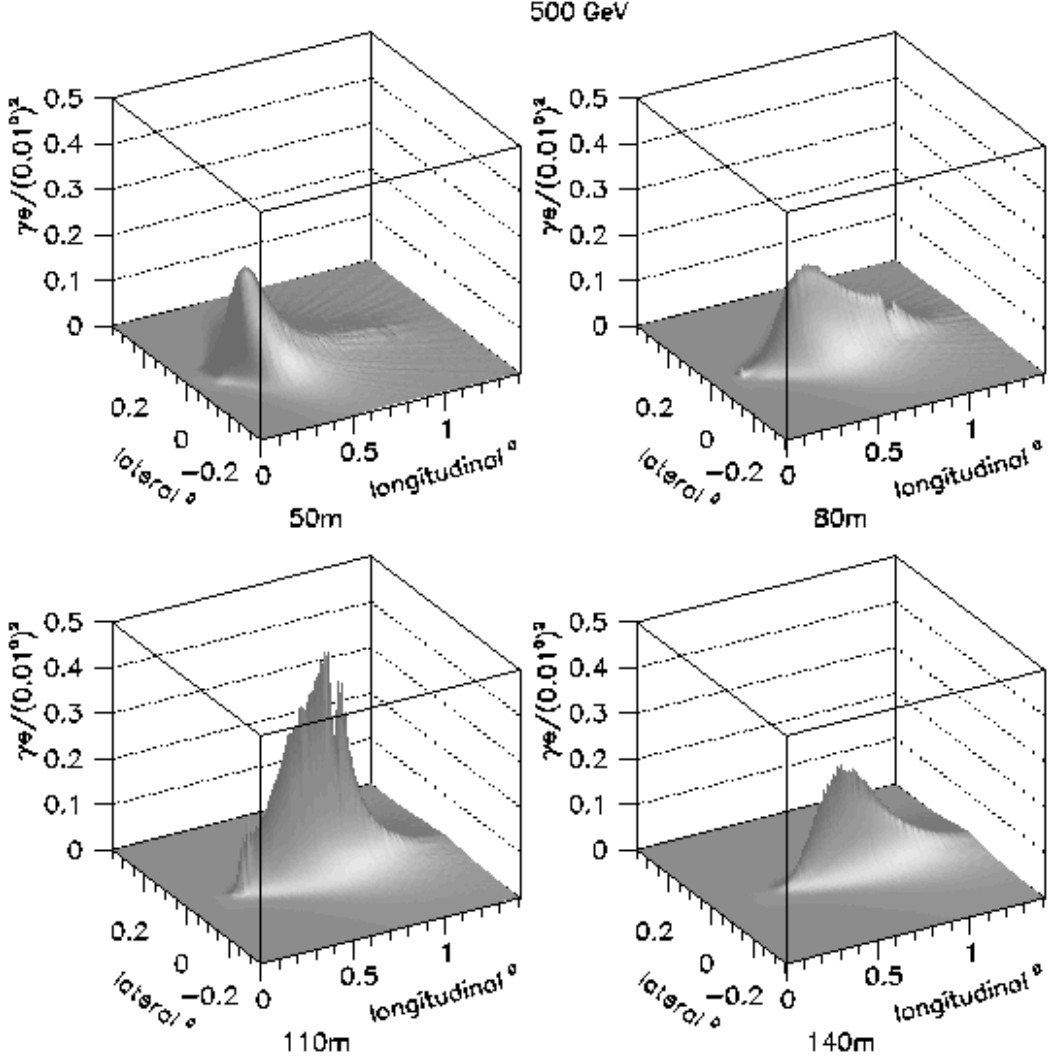}
\caption{Bidimensional image profiles of $500\:{\mathrm {GeV}}$ vertical
$\gamma$-showers ($\gamma$e density) for various values of impact
parameter as given by the semi-analytical model. The $x$ (longitudinal)
and $y$ (lateral) axes are coordinates in the focal plane in degrees. 
The source is at the origin.  Note that the lateral scale is dilated by
a factor of two.}
\label{fig:model}
\end{figure}

\begin{figure}
\epsfxsize=14.5 cm
\leavevmode
\centering
\epsffile[5 25 540 240]{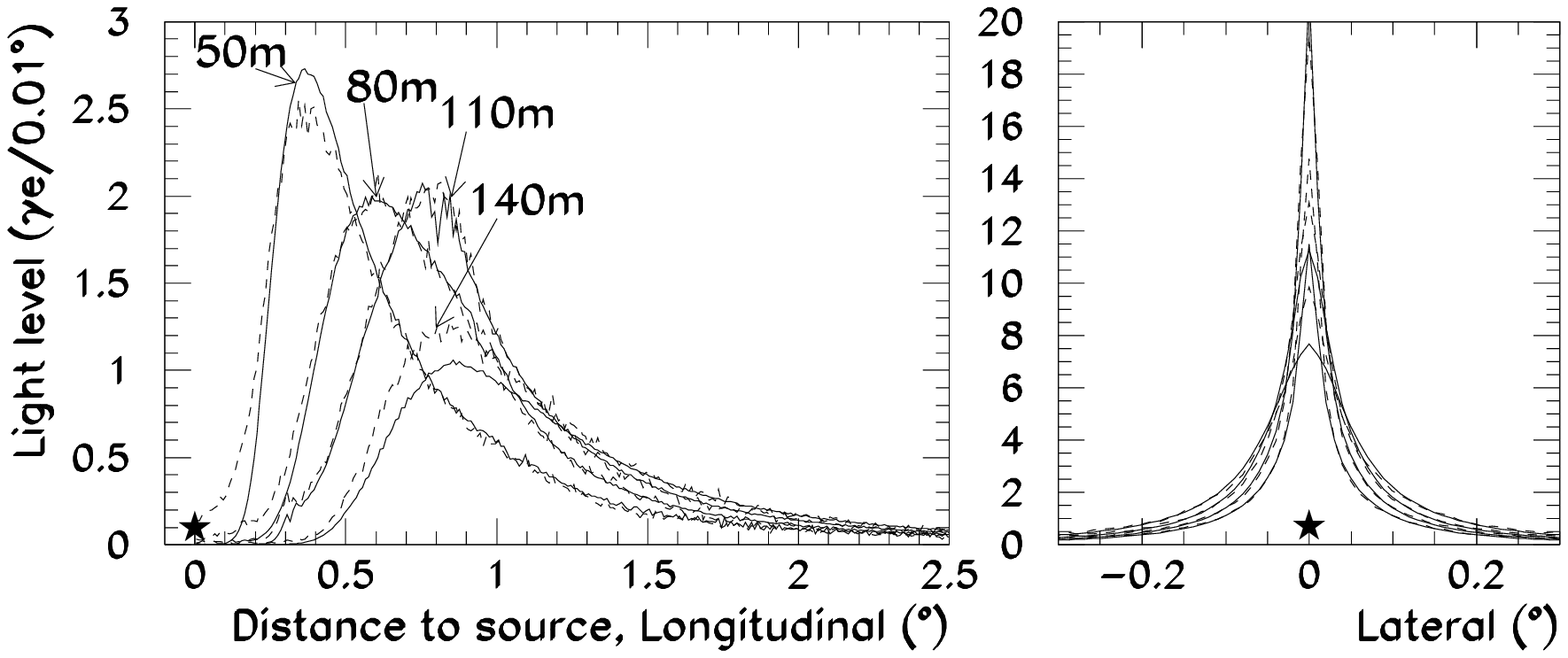}
\caption{Longitudinal and lateral image profiles ($\gamma$e density
projected on the lateral/longitudinal axis) of $500\:{\mathrm {GeV}}$
vertical $\gamma$-showers for various values of impact parameter
($D$).  The source is at the position of the star.   The image profiles
as given by the semi-analytical model are shown by the full lines, the
average image profiles from full Monte  Carlo simulations by dotted
lines.  For the lateral profiles, the width of the profile decreases
with increasing $D$. }
\label{fig:latlon}
\end{figure}

\begin{figure}
\epsfxsize=14.5 cm
\leavevmode
\centering
\epsffile[10 170 600 660]{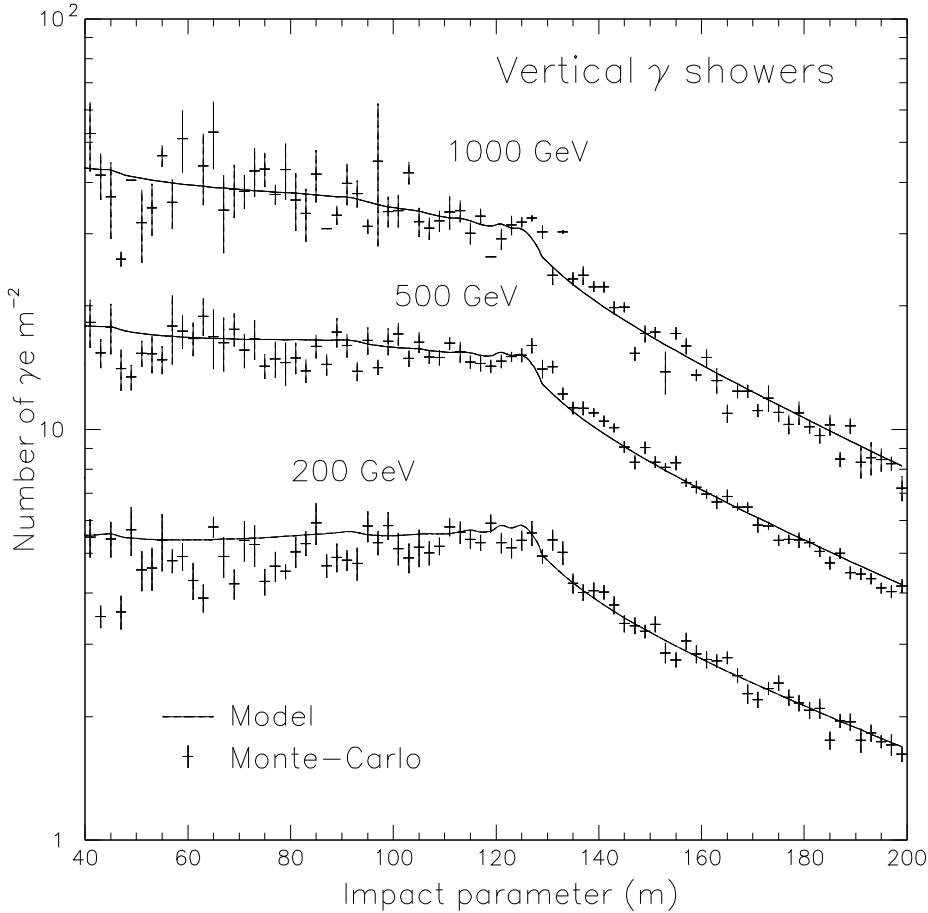}
\caption{Density of detected photons on the ground as a function of the
impact parameter: the full line corresponds to the semi-analytic model
described in the text; points with error bars show the result of full
Monte Carlo simulations.}
\label{fig:densite}
\end{figure}

In practice, the shower (for a given energy and impact parameter) is
divided into ``slices'' perpendicular to its axis at different depths,
and the contribution of each slice to the image is calculated. In such
a slice, the contribution of a charged particle ($e^{\pm}$) with a
given energy and direction and a given lateral position with respect to
the shower axis is evaluated as follows: the corresponding cone of
Cherenkov emission traces a circle on the plane perpendicular to the
shower axis containing the imaging telescope; rather than considering a
mirror with a fixed position with respect to the shower, we allow it to
rotate around the shower axis (the distance  $D$ is fixed) and use the
intersection of the corresponding ring with the circle of Cherenkov
light to find  the average contribution of these $e^{\pm}$'s to the
image as well as the position of the photo-electrons in the focal plane
(see Appendix~2 and Fig.~\ref{fig:calc}).  Summing over the energy,
direction, and averaging over the lateral position of all $e^{\pm}$'s
in the slice gives the mean position  of its image in the focal plane,
its mean transverse extension, and the corresponding density of
$\gamma$e's. The sum of the contributions of all the slices gives a
full mean image description in terms of the linear density of light
along the image axis and the transverse extension of the image as a
function of position along this axis. The form of the transverse
profile is taken to be a constant shape, determined from Monte-Carlo
simulations.  This form is scaled to the calculated transverse
extension, which varies with position along the image axis.  The
bi-dimensional profile of vertical $500\:{\mathrm {GeV}}$ $\gamma$-ray
showers thus obtained is shown in Fig.~\ref{fig:model} for different
values of the impact parameter $D$.  
It can be seen that the longitudinal profiles
of  the expected images are asymmetrical and that their position and 
general shape depend on $D$. The consistency between this analytical
calculation and the full Monte-Carlo simulation can be checked in  
Fig.~\ref{fig:latlon}, which shows the lateral and  longitudinal
profiles of the $\gamma$-images for various impact  parameters, both
from full Monte Carlo simulations and as given by the model. 
It can
also be seen in Fig.~\ref{fig:densite}, which shows the mean density
of Cherenkov light on the ground in a $4.6^\circ$ acceptance
calculated by both methods, and  found to be in good agreement.

For the comparison with an individual event image, the expected
distribution of $\gamma$e's in the focal plane (for given values of
$E_{\gamma}$, $D$, and $\phi$ of the shower axis projected in the image
plane) is integrated over each pixel area, according to the source
angular position (defined by two angular coordinates $\vec{\xi}$). A
function $\chi^{2} (E_{\gamma}, D, \vec{\xi}, \phi)$ is then defined in
such a way that  its minimization with respect to $E_{\gamma}$, $D$,
$\vec{\xi}$, and $\phi$ gives the values of those parameters which best
fit the model to the image. This function is a sum of squared
differences between expected and actual pixel contents divided by a
quadratic error, extending over all the pixels whose expected or actual
content is above a given threshold ($2\:\gamma$e). In the following
expression, $Q^{\mathrm{real}}_i$ is the actual value of the charge 
collected in pixel~$i$ (measured in equivalent number of
photo-electrons), $Q^{\mathrm{mean}}_i$ is the expected mean
contribution of the shower  to pixel~$i$ for a given set of model
parameters, and $\overline{B_i}$ is the mean contribution of the noise
calculated on the basis of measured randomly-gated events, which 
contributes to the fluctuations but not to the charge, due to the
{\small AC} coupling.  For the C{\small AT} detector, the electronic
noise  dominates the night-sky background, since the mirror and pixel
size are small and the gate is short. The fluctuations in
$Q^{\mathrm{real}}_i$ and $Q^{\mathrm{mean}}_i$ are supposed 
proportional to the square root of the charge in the pixel, giving:
\[ 
\chi^{2} = \frac{1}{k} \sum_{i} 
\frac{(Q^{\mathrm{real}}_{i} - Q^{\mathrm{mean}}_{i})^{2}} 
{\overline{B_i} + \frac{1}{2} 
(Q^{\mathrm{real}}_{i}+Q^{\mathrm{mean}}_{i})}  
\]
The preceding expression is not a $\chi^{2}$ {\it stricto sensu} since
the exact number of degrees of  freedom depends somewhat on the value
of the fitted parameters through the number of expected hit
photo-tubes. The value of the error factor~$k$ (here taken as 2.9) has
no effect on the value of the fitted shower parameters. It has been
adjusted on simulations in order to have an approximately flat
$\chi^{2}$ probability for simulated $\gamma$-shower images well above
the trigger threshold. The use of an energy or impact parameter
dependent $k$-value is under study. In the case of well-localized point
sources, fits in which $\vec{\xi}$ is fixed are used for improved
energy resolution.

The fit is quite sensitive to the starting values, so a good estimation
of these values is needed.  For this purpose the usual moment-based
parameters are calculated, but after a ``principal cluster''
image-cleaning procedure.  In this procedure the pixels with a charge
less than a threshold of $2\:\gamma$e are zeroed; the principal cluster
consists of the largest number of contiguous  pixels, all others are
then zeroed.  This provides moment-based parameters for $\gamma$-images
which are more stable against the noise background (though these would
not be useful for background rejection, as this cleaning procedure
makes the hadronic background more ``$\gamma$-like'').  Relations
between the moment-based parameters and the values to be fitted have
been defined on Monte-Carlo simulations; these are then used to define
the starting values for the fit.

\section{Results of the method when applied to the CAT images}

Many of the sources in the E{\small GRET} catalogue \cite{egret} have 
well-identified radio, optical, or X-ray counterparts, which give the
source position to an accuracy much better than can be achieved by
{\small ACT}  telescopes.  Such sources with known position are usually
placed  at the centre of the field of an imaging telescope.  However,
many unidentified sources in the  E{\small GRET} catalogue are located 
in error boxes with typical size of 1$^\circ$.  Ground-based  Cherenkov
detectors are, in principle, able to localize such sources  with a
higher precision.  In order to observe these sources,  different
methods have been developed by imaging Cherenkov telescope  groups. 
Stereoscopy is the most direct way to find the direction  of a source,
but requires at least two telescopes and reduces  the collection area
\cite{hegratel}.  For single-telescope experiments with sufficient 
background rejection on an image-shape criterion, it is possible  to
perform de-localized analyses assuming the source at the different 
points of a grid covering the field of view or by examining the
distribution of the intersections  of the image axes \cite{grid}. 
However, an event-by-event analysis method  such as that described here
is  preferable since in the former methods the signal is more easily 
drowned-out by the background.  

The method has been tested on simulated  $\gamma$-images provided by
the Monte-Carlo simulation  program described above and a realistic
simulation of the detector response, including the measured variation
in collection efficiency and gain  between the phototubes, and measured
wavelength response of the mirrors and Winston cones. For optimization
of the cut values, these simulated $\gamma$-images have been used
together with the real background  events from data from off-source
runs with the C{\small AT} imaging telescope. Gammas from a point-like
source with the Crab nebula spectrum \cite{catcrab} (see equation (1)
above) were simulated at various elevations.  The capability of the
method both  for source detection and for source spectrum measurement
have been examined. 

\subsection{Source detection}

\begin{figure}
\epsfxsize=14.5 cm
\leavevmode
\centering
\epsffile[0 25 590 520]{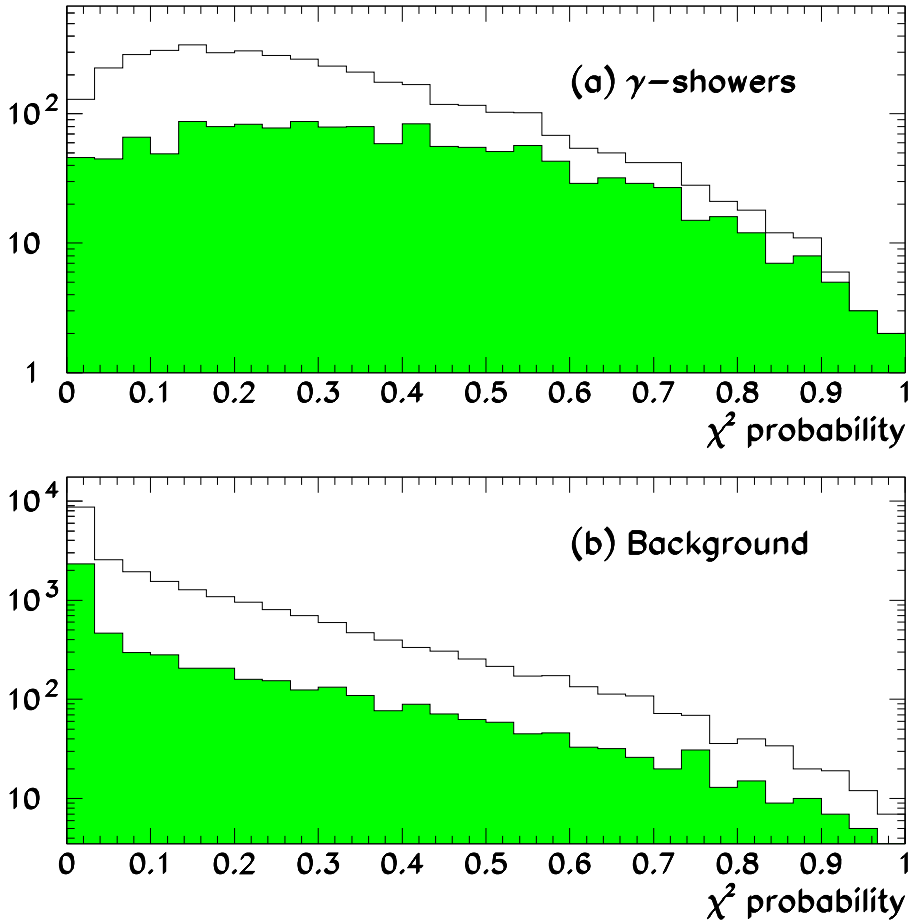}
\caption{$\chi^2$ probability distribution for a fit with the source
coordinates considered as free parameters (constraint is from shape
alone):
a) vertical $\gamma$-ray showers; 
b) real background (off-source data) showers.
The upper line in each figure is for all events above threshold, the
shaded distributions for events with a fitted energy, $E_{\mathrm{f}}$,
greater than $350\:{\mathrm {GeV}}$ and a fitted impact parameter,
$D_{\mathrm{f}}$, between 30 and $125\:{\mathrm{m}}$.}
\label{fig:chi2}
\end{figure}

\begin{figure}
\epsfxsize=14.5 cm
\leavevmode
\centering
\epsffile[20 25 540 230]{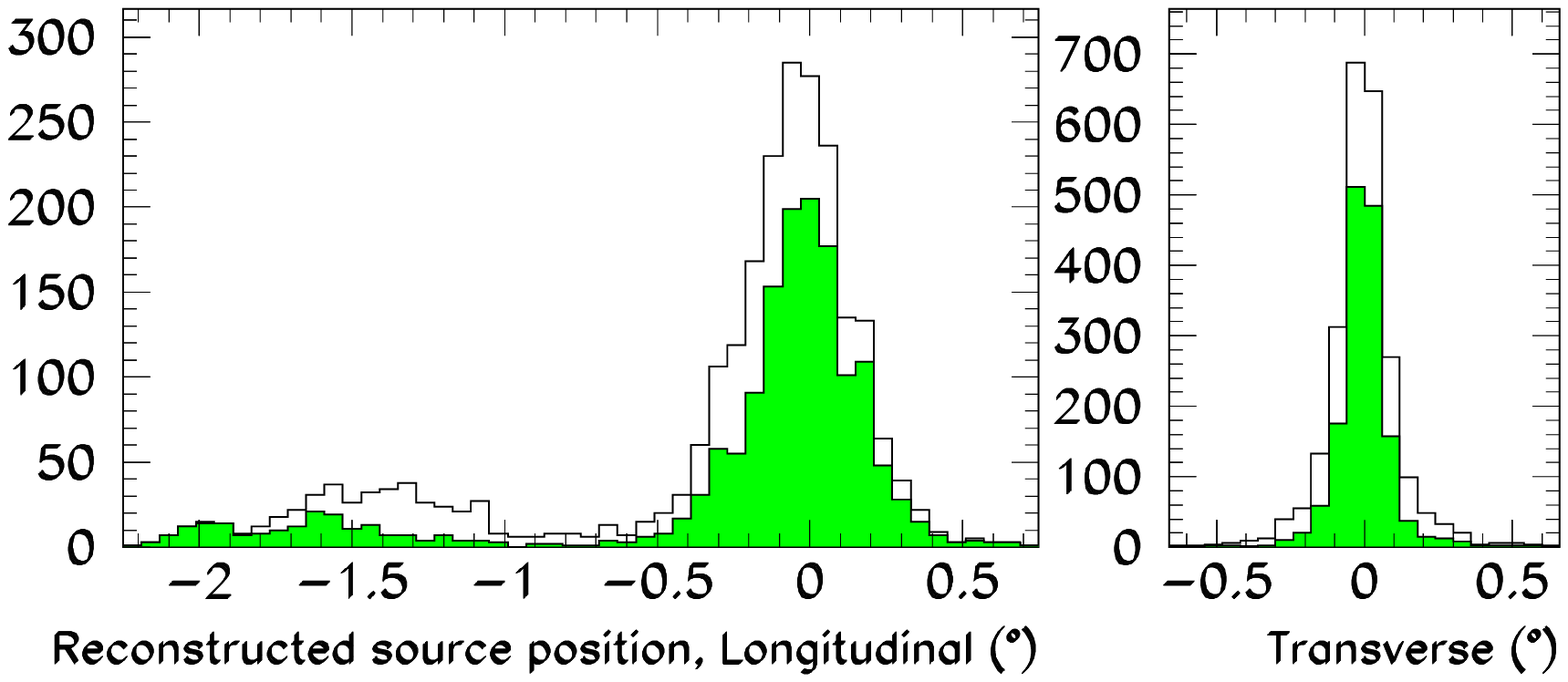}
\caption{Distributions of longitudinal and transverse errors (with
respect to the image major axis) in the reconstruction of the source
position for vertical showers with $P(\chi^2)>0.2$.  The small bump at
negative values of longitudinal error results from wrong direction
reconstruction (mainly for events close to the energy threshold). 
Shaded histograms correspond to events with a fitted energy greater
than $350\:{\mathrm GeV}$.} 
\label{fig:pointerr}
\end{figure}

\begin{figure}
\epsfxsize=14.5 cm
\leavevmode
\centering
\epsffile[0 25 580 520]{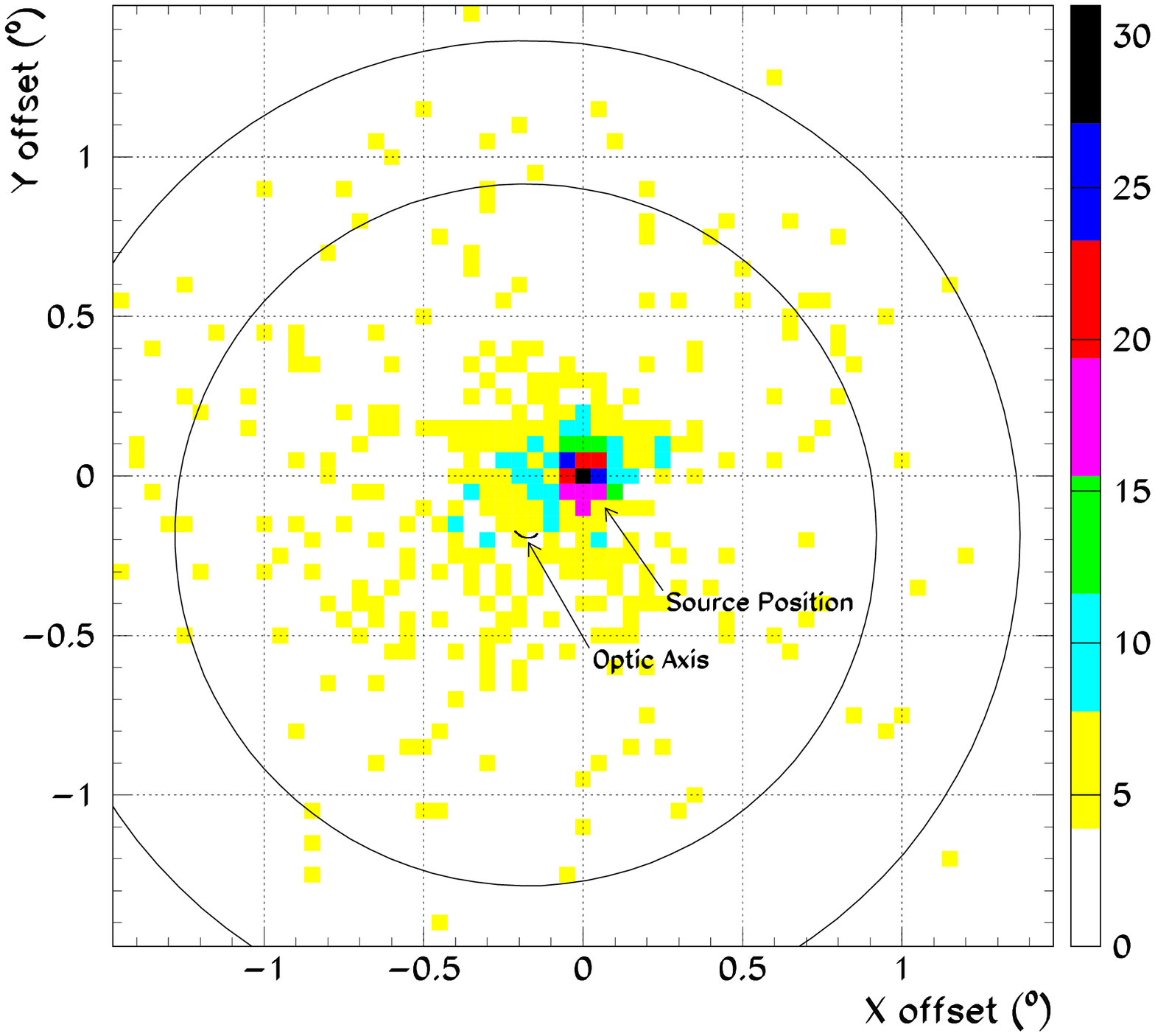}
\caption{The distribution of reconstructed angular origins for the data
from a 30-minute run on Markarian 501 from April 16, 1997, for events
with $P(\chi^2)>0.2$. The concentric circles represent the trigger
region and the small-pixel region, respectively.  During the run, the
optic axis described the small arc indicated due to the mechanical
flexibility of the structure, which is monitored as  described in [9].
The number of events reconstructed in each bin of $(0.05{^0})^2$ is
shown.  No  background subtraction has been performed.} 
\label{fig:m501}
\end{figure}

As applied to the data, the method consists of minimizing the  $\chi^2$
with respect to $E_\gamma$, $D$, $\vec\xi$, and $\phi$. 
Fig~\ref{fig:chi2} shows the $\chi^2$ probability distributions
obtained  with this fit for the simulated $\gamma$-ray events and real
background events.  
A cut on the $\chi^2$ probability value, $P(\chi^2)$, provides a
selection of $\gamma$-like events on the basis of the image shape
alone. The reconstructed angular  origins obtained for simulated
$\gamma$-events accumulate around the actual source position which in
this case is at the centre of the field.  The dispersion around the
actual source position  has a typical {\small RMS} spread of
$0.14^\circ$.  For each event, the accuracy of the angular origin
determination is better by a factor two in the direction perpendicular
to the image axis than in the direction of the image axis 
(Fig.~\ref{fig:pointerr}).  
The {\small RMS} longitudinal error typically varies  from $0.2^\circ$
to $0.1^\circ$ as the energy varies from the threshold to $2\:{\mathrm
{TeV}}$.  The angular origins obtained for background events are spread
over the whole field with an approximately Gaussian distribution with a
$1.8^\circ$ {\small FWHM}. Since this distribution is fairly flat,
2-dimensional skymaps of the  angular origins of the showers could be
used for source detection, as can be seen from the reconstructed
positions in data taken on Markarian 501 in Fig.~\ref{fig:m501}.  
The errors in angular reconstruction given above are for a single
shower; a point source with poorly defined position could thus be
localized to $\sim 1-2'$ with the combination of $\sim 100$ such
events.

\begin{figure}
\epsfxsize=14.5 cm
\leavevmode
\centering
\epsffile[5 25 580 520]{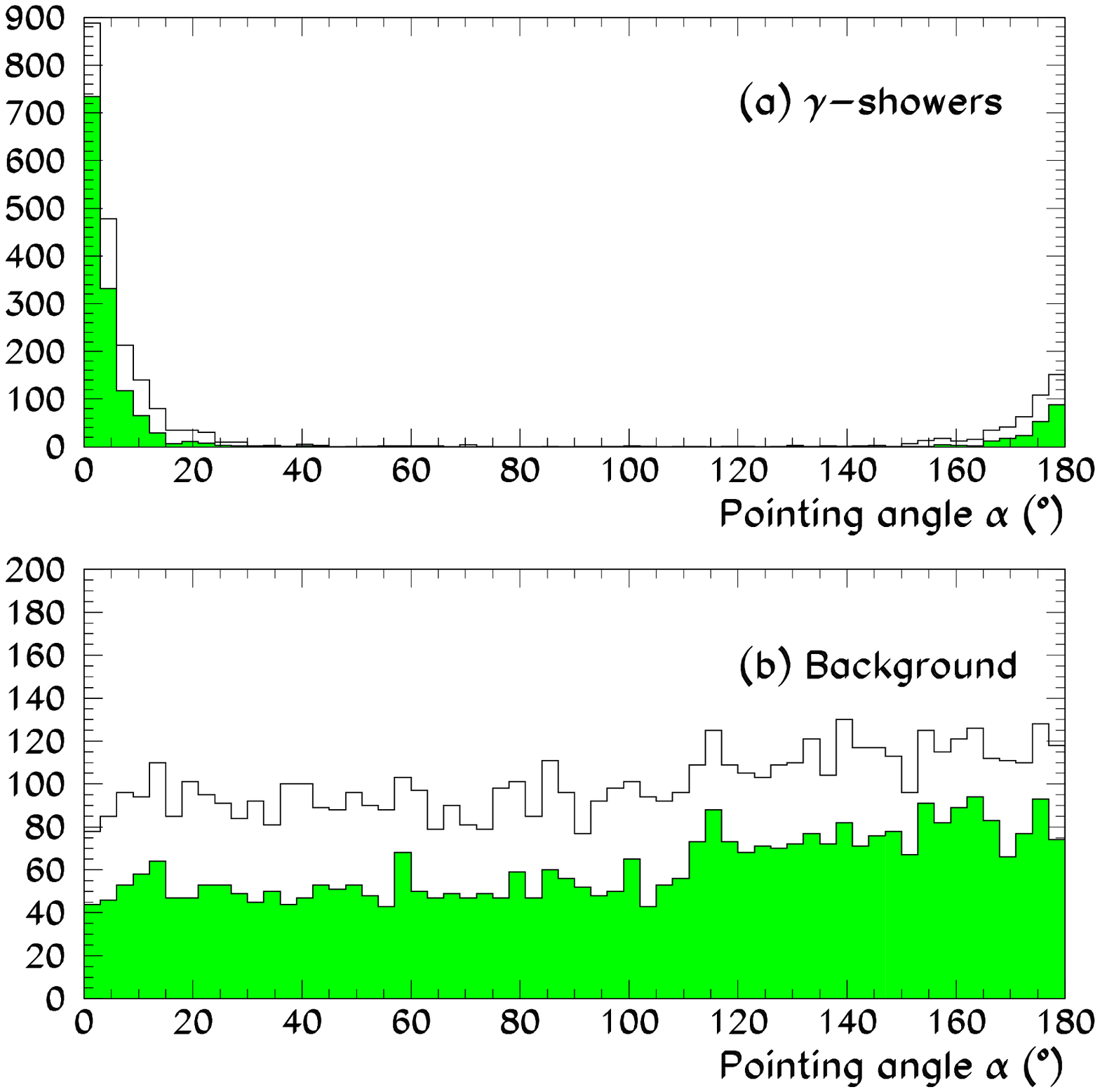}
\caption{Distribution of the pointing angle $\alpha$ for events with
$P(\chi^2)>0.2$ (constraint from shape alone):
a) vertical $\gamma$-ray showers; 
b) Real background (off-source data) showers.
The upper line is for all events above threshold, the shaded histograms
for events with a fitted energy greater than $350\:{\mathrm {GeV}}$.}
\label{fig:alpha}
\end{figure}

For the present, a conservative procedure of monitoring the background
is used,  based on the pointing angle $\alpha$, similar to the
``orientation'' of Whipple  \cite{scuts}: $\alpha$ is the angle at the
image barycentre between the actual  source position and the
reconstructed source position. The pointing angle does  not use the
full information contained in the results of the fit, but has a  fairly
flat distribution from $0^\circ$ to about $120^\circ$ for background 
events, which allows the background level to be easily monitored.   The
cut on $\alpha$ is more efficient than a cut on the angular distance
between the source position and the reconstructed $\gamma$ origin
since, as seen in  Fig.~\ref{fig:pointerr}, the position
reconstructed is not symmetric about the source position. The
distribution of $\alpha$ for $\gamma$-events from a Crab-like source
exhibits a peak at $0^\circ$ (Fig.~\ref{fig:alpha}) and a small
accumulation at $180^\circ$ corresponding to events which are
wrongly found to point away from the source.
Around 17\% of the
$\gamma$-events from a Crab-like source are in this situation.  The
proportion of events with a wrongly reconstructed direction decreases
with increasing energy, from 22\% at $200\:{\mathrm {GeV}}$ to 9\% at 
$600\:{\mathrm {GeV}}$ and 4\% at $1\:{\mathrm {TeV}}$.   

\begin{figure}[t]
$$
\epsfxsize=6.5 cm
(a)\epsffile[40 30 540 490]{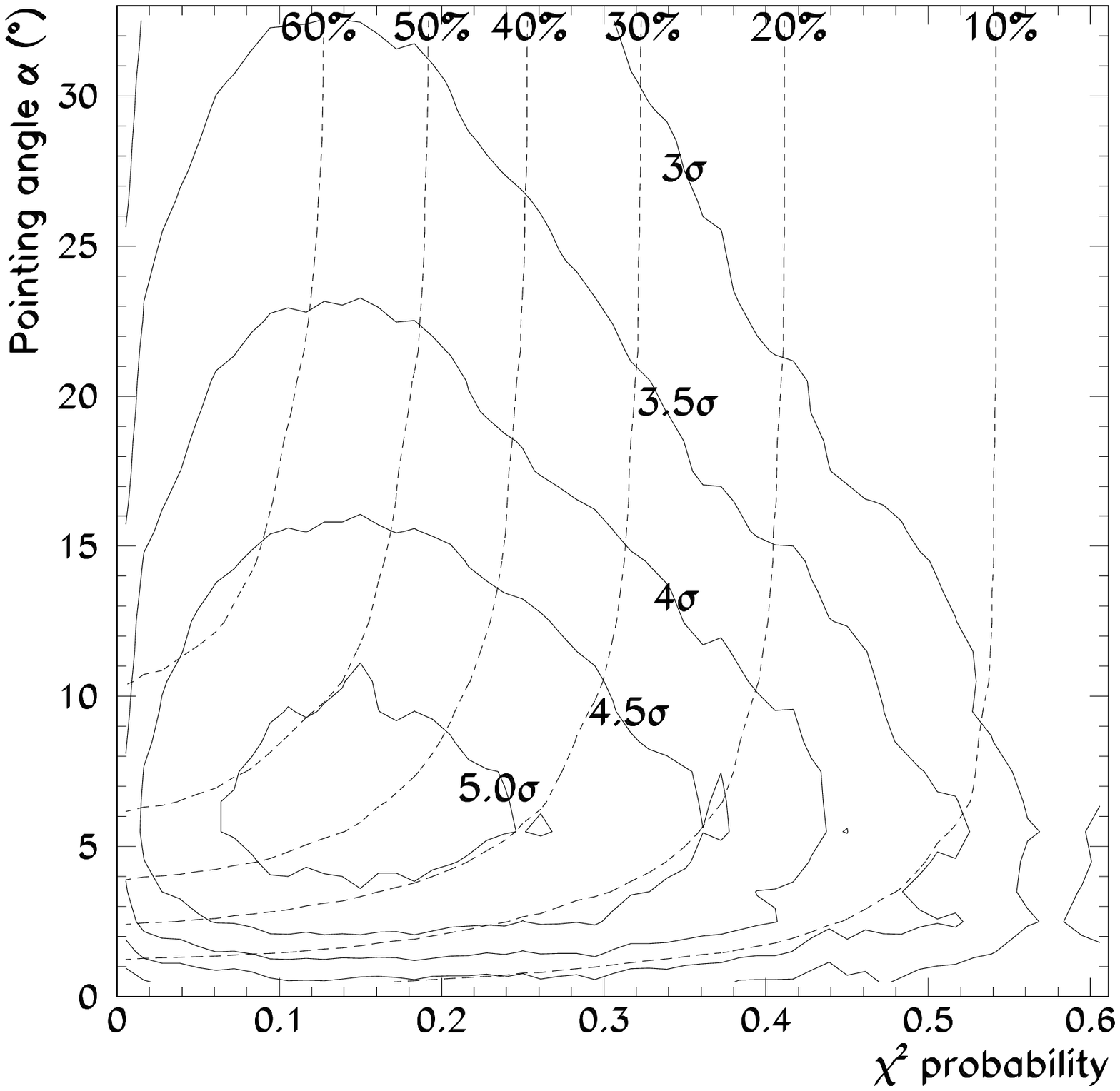}
\;
\epsfxsize=6.5 cm
(b)\epsffile[40 30 540 490]{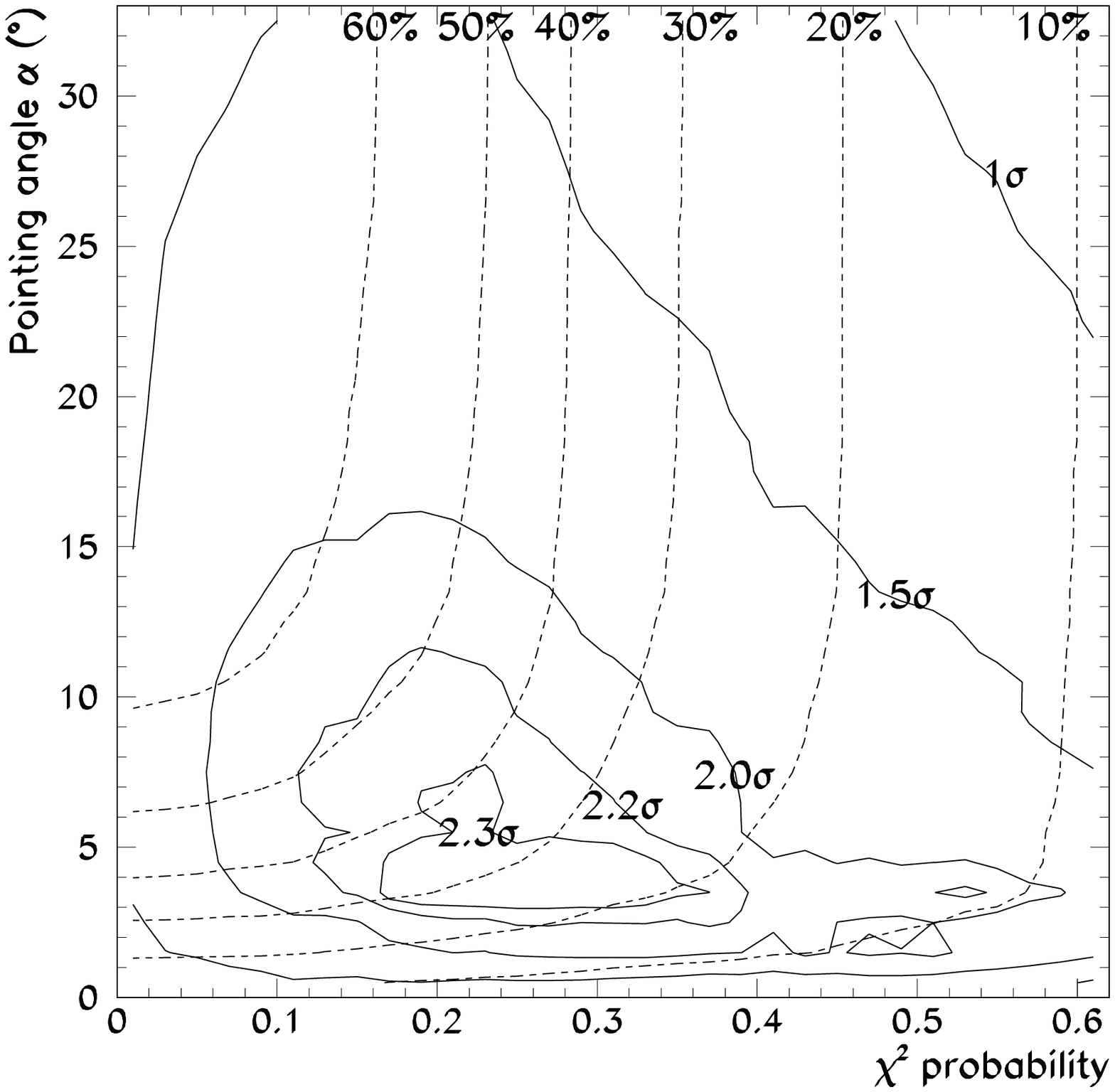}
$$
\vspace{-1cm}
\caption{Detection significance for a Crab-like source 
at zenith in one hour of observation, shown as a function of
the two cuts on $P(\chi^2)$ and $\alpha$:
(a) for a source at the centre of the camera;
(b) for a source at $1^\circ$ from the centre.
The full lines indicate contours of equal significance; dotted
lines show fixed $\gamma$-ray selection efficiency.}
\label{fig:signif1}
\end{figure}

The significance of a signal is calculated using the usual formula: 
$(ON-OFF)/\sqrt{ON+OFF}$ \cite{lima}, assuming equal time on and
off-source.  The significance per hour on a simulated Crab-like source
at zenith has been calculated for various cut values on $\alpha$ and
$P(\chi^2)$ (Fig.~\ref{fig:signif1}.a).  
The best result in terms of both significance and efficiency for
$\gamma$-events is obtained for a $P(\chi^2)>0.2$ and $\alpha<6^\circ$,
which gives $5.1\sqrt{t}\ \sigma$ where $t$ is the on-source 
observation time in hours, retaining 34\% of the $\gamma$-events  while
giving a rejection factor of 120 on background events.  This rejection
factor is smaller than for some comparable  experiments as there is a
large rejection factor at the trigger level, allowing a moderate
background rate of $15\:{\mathrm {Hz}}$ at the zenith. At $45^\circ$
from zenith the best significance for the same cuts falls to
$2.7\sqrt{t}\ \sigma$. This is essentially due to the higher energy
threshold, leading to a lower event rate; on the other hand, for the
same $\gamma$-ray selection efficiency the background rejection factor
is comparable to that at zenith.

In order to estimate the efficiency of the $\chi^2$-method in the case
of a source with a poorly-defined position, a simulated Crab-like
source has been set on the edge of the trigger area ($1^\circ$ from the
centre). In this case, the equivalent detection area is divided by a
factor of the order of two.  Even for a source with known position,
when the source is not at the centre of the field it is possible to use
one side of the camera as the off region for the other side
\cite{offc}.  For a Crab-like source at zenith, the same cuts in
$P(\chi^2)$ and $\alpha$ as for a source in the centre of the camera
give a $2.3\sqrt{t}\ \sigma$ significance, with a selection efficiency
for $\gamma$'s of 36\% and a rejection factor of 116 above threshold
(Fig.~\ref{fig:signif1}.b). This means that the on-source run time
has to be five times larger for a source on the edge of the trigger
area than for a source at the centre of the field for the same
significance.  The corresponding significance obtained at $45^\circ$
from zenith for an  off-centre source is $1.5\sqrt{t}\ \sigma$.

\subsection{$\gamma$-ray energy measurement}

\begin{figure}
\epsfxsize=14.5 cm
\leavevmode
\centering
\epsffile[15 25 550 300]{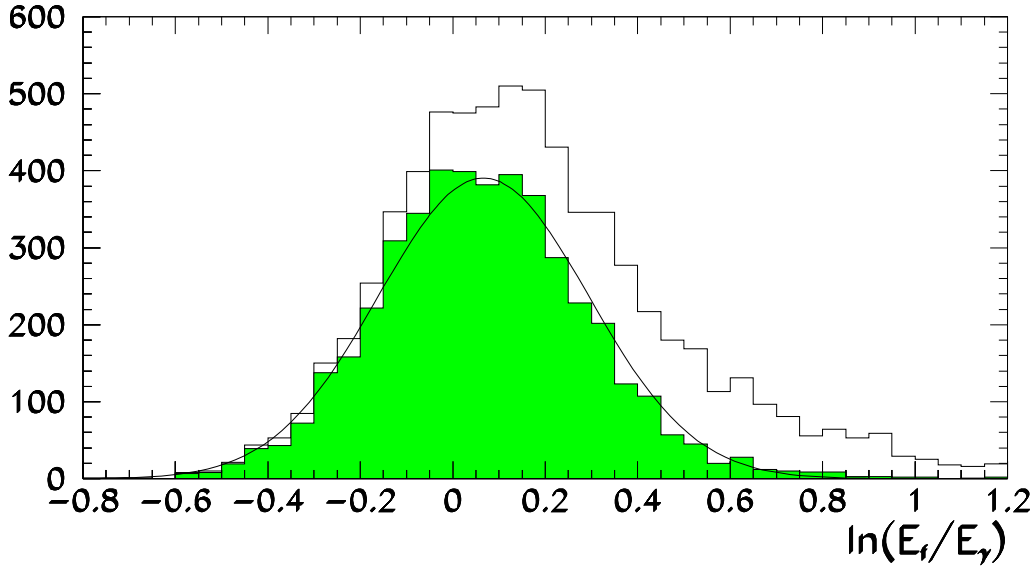}
\caption{Distribution of $\ln({E_{\mathrm {f}}}/{E_\gamma})$ for
vertical $400\:{\mathrm {GeV}}$ $\gamma$-ray showers satisfying the
selection cuts (the source location, $\vec{\xi}$, being fixed in the
fit). The shaded histogram is further restricted to events with a
fitted impact parameter $30\:{\mathrm {m}} < D_{\mathrm {f}} <
125\:{\mathrm {m}}$.}
\label{fig:ener500}
\end{figure}

\begin{figure}
\epsfxsize=14.5 cm
\leavevmode
\centering
\epsffile[5 25 580 520]{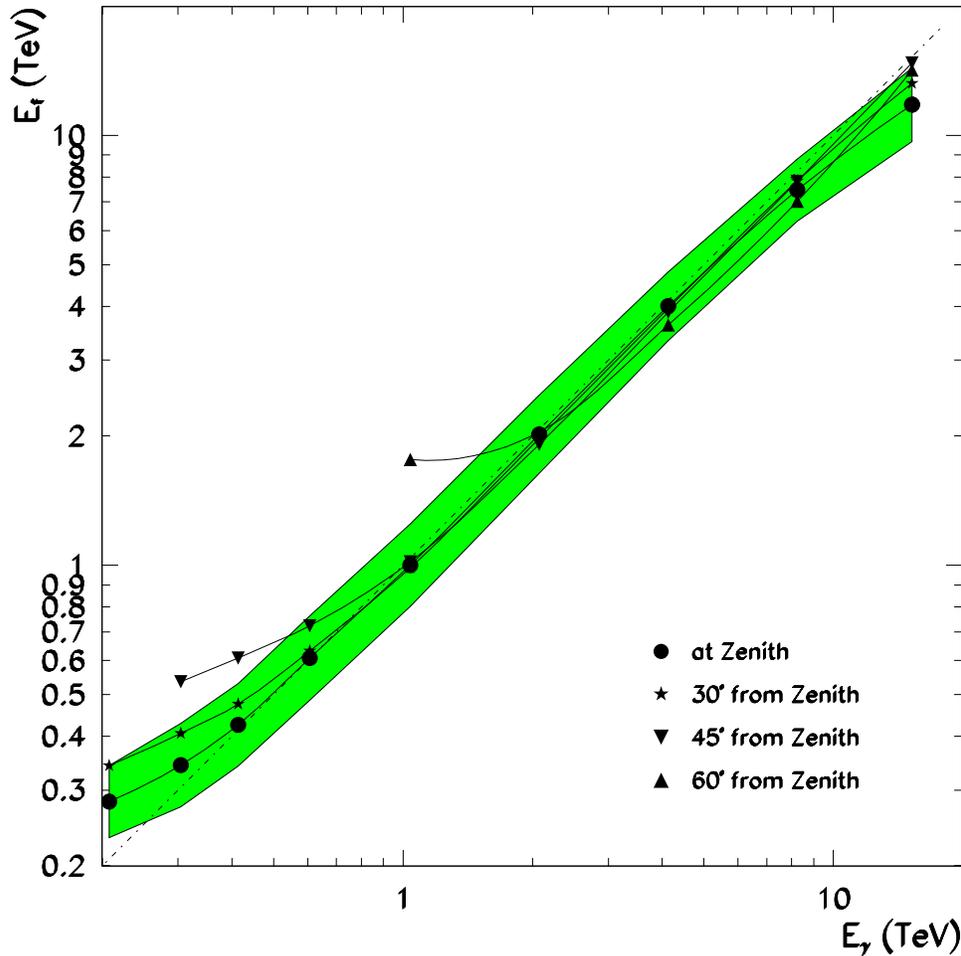}
\caption{Fitted energy, $E_{\mathrm {f}}$, versus true energy, $E_\gamma$,
for $\gamma$-ray showers satisfying the selection cuts (the source
location, $\vec{\xi}$, being fixed in the fit) and for which
$30\:{\mathrm {m}}< D_{\mathrm {f}}\cos Z < 125\:{\mathrm {m}}$
($Z=$~zenith angle). The shaded interval shows the 68\%  confidence
intervals for a source at the zenith.  The effect of trigger selection
on the energy estimation can be seen.}
\label{fig:enerfit}
\end{figure}

For a source detected with  a strong enough significance, the energy
spectrum can be studied by a detector with good energy resolution.  The
fit described above in which the source position is a free parameter
gives a first estimate of the energy of each event to within about
30\%. However, more precise spectral studies can be carried out on
point sources of $\gamma$-rays. The use of the source position as a
constraint in the fit provides a higher accuracy for impact parameter
measurement and, as a consequence, for energy measurement. If trigger
selection effects are ignored in the Monte-Carlo program (thus
accepting all events above $100\:{\mathrm {GeV}}$), the $\chi^2$
minimization with respect to $E_\gamma$, $D$, and $\phi$  provides an
unbiased energy measurement within about 25\% (statistical error only).
Trigger selection effects are small for events well above the
threshold, as can be seen for simulated $400\:{\mathrm {GeV}}$
$\gamma$-rays in Fig.~\ref{fig:ener500}.
This figure also shows that the distribution of the fitted event
energies about the true energy is Gaussian on a logarithmic scale. 
Consequently, the slope of a power-law spectrum can be directly
estimated with this technique. Close to the threshold, however, the
fitted energy $E_{\mathrm {f}}$ is overestimated as a consequence of
the trigger selection. Similarly, the small remaining bias in $ \log
(E_{\mathrm {f}}/E_\gamma)$ at $400\:{\mathrm {GeV}}$ is due to
showers with large impact parameters for which the trigger selection is
critical at this energy since the telescope is then located at the
border of the light pool (Fig.~\ref{fig:densite}). This effect is
largely removed if only showers with a fitted  impact parameter
$D_{\mathrm {f}}$ lower than $125\:{\mathrm {m}}$ are included (shaded
histogram in Fig.~\ref{fig:ener500}). The bias induced by the trigger
selection at different energies is best illustrated  by plotting 68\%
confidence intervals for $E_{\mathrm {f}}$ as a function of the true
value $E_\gamma$ used in the simulation (Fig.~\ref{fig:enerfit}).
It can be seen that for $E_f$ below $350\:{\mathrm {GeV}}$ for vertical
showers, only an upper limit can be safely derived for $E_\gamma$.
Therefore, spectrum measurement is reliable only above a spectrometric
threshold, that is, in the region in which $E_{\mathrm {f}}$ depends
linearly on $E$. It is somewhat higher than the nominal threshold which
is relevant for source discovery. Fig.~\ref{fig:enerfit} also shows
the variation of the average  values of $\log(E_{\mathrm {f}})$ on
$\log(E_\gamma)$ for zenith angles $0^{\circ}$,  $30^{\circ}$,
$45^{\circ}$, and $60^{\circ}$, showing the increase in the 
spectrometric thresholds with increasing zenith angle. Restricting to
events with $E_{\mathrm {f}}$ above the spectrometric threshold and
$30\:{\mathrm {m}}<  D_{\mathrm {f}} \cos Z <  125\:{\mathrm {m}}$, 
the accuracy of the preceding method is about 20\% (statistical error
only), independent of the zenith angle $Z$ up to $45^{\circ}$.  


\section{Conclusion}

The method described above, based on a realistic analytic description
of electromagnetic air showers, is best suited for those Cherenkov
Imaging Telescopes with a high-resolution camera. The light
distribution in the focal plane is fully exploited, yielding the shower
direction from the asymmetry of the longitudinal profile as well as the
source position in the focal plane. Selection of $\gamma$-rays on the
basis of the image shape is performed by using a single
$\chi^2$-variable instead of a series of cuts on various image
parameters.  By combining the $\chi^2$ probability cut and a direction
($\alpha$) cut, a significance of 5$\sigma$ per hour can be achieved
for a Crab-like source at the zenith. A future development of the
method would be to use the distribution of selected $\gamma$-ray
origins on the celestial sphere together with the known
energy-dependent point spread function of the method to estimate the
significance by a maximum-likelihood method. With the C{\small AT}
telescope ($250\:{\mathrm {GeV}}$ threshold), sources with the
intensity of the Crab nebula can be  detected in one hour.  This has
been confirmed with the results obtained in the  96/97  observing
campaign.  Moreover, sources with poorly defined position can be 
localized with an accuracy of the order of an arc minute on the basis
of about  100 showers. The accuracy on $\gamma$-ray energy is of the
order of 20--25\%. Biases induced by the trigger selection have been
investigated; in particular, care should be taken in spectrum
measurement, which is accurate only above a specific threshold, higher
than that used for source detection.

\section{Appendix 1 : Angular distribution of $e^{\pm}$'s in the
analytic model}

Let $\theta_{\mathrm {e}}$ be the angle of a charged particle from the
shower axis and $E_{\mathrm {e}}$ its kinetic energy.
In \cite{hillas}, the variable of interest is
\[
w = \frac{2(1 - \cos \theta_{\mathrm {e}})}
{(E_{\mathrm{e}}/21\:{\mathrm{MeV}})^2} \;\;\;.
\]
In order to be in good agreement with the Monte-Carlo program
\cite{whimc}, the average value of $w$ has been parametrized as 
\[
\overline w = \frac{0.9}{1 + 120\:{\mathrm{MeV}}/E_{\mathrm {e}}}
\]
instead of the corresponding formula given in \cite{hillas}.

\section{Appendix 2 : Calculation of the mean image longitudinal and
transverse profiles}

\begin{figure}
\epsfxsize=10.2 cm
\leavevmode
\centering
\epsffile[0 0 350 730]{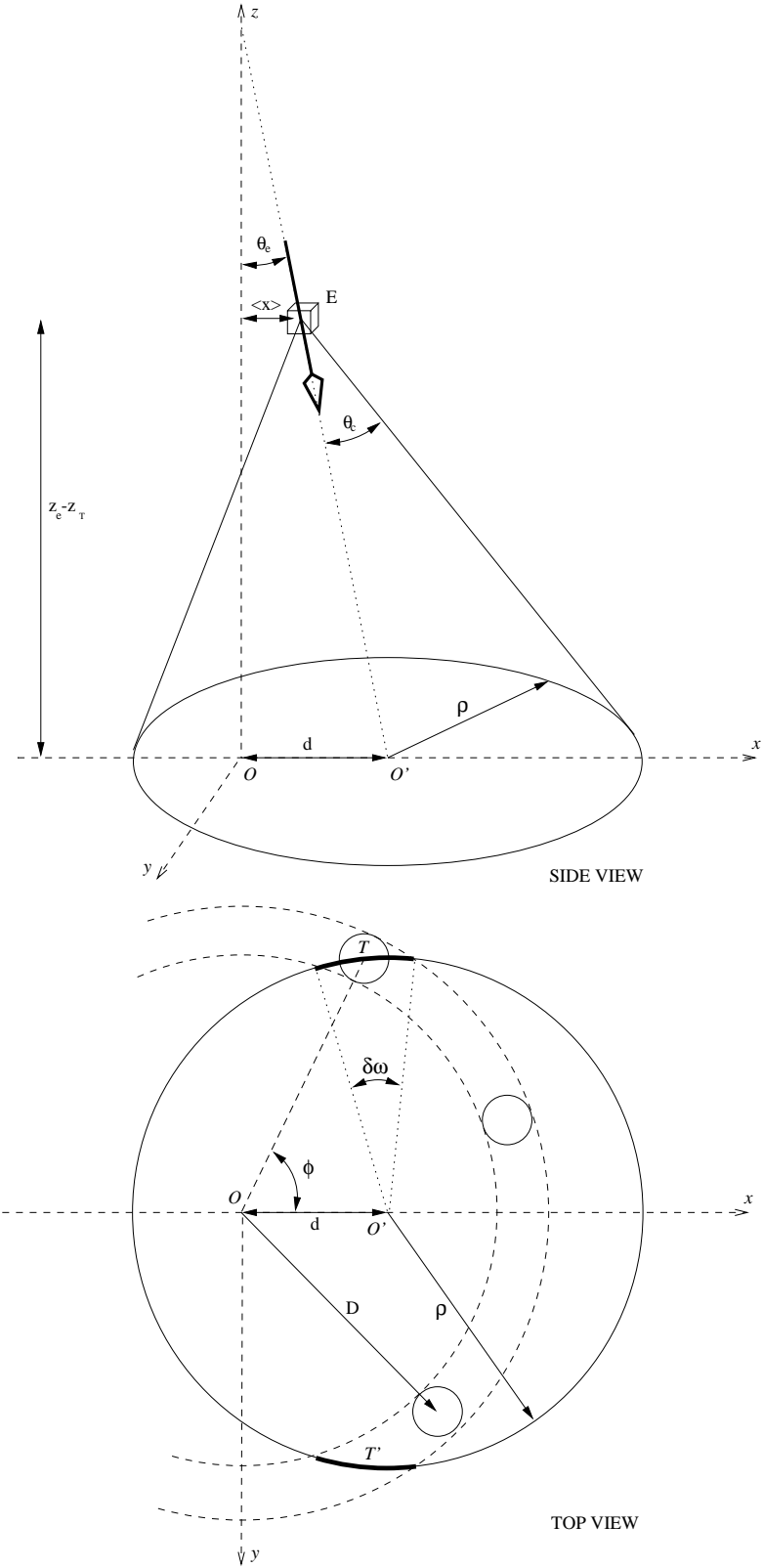}
\caption{Geometric construction used in the derivation of the
semi-analytical model.  See Appendix 2 for explanation.}
\label{fig:calc}
\end{figure}

The notation used in the calculation of the semi-analytical model is
shown in Fig.~\ref{fig:calc}. The origin $O$ is taken on the shower 
axis $Oz$ as the point closest to the telescope mirror.  We consider a
charged particle ($e^{\pm}$) at an angle $\theta_{\mathrm {e}}$ from
the  shower axis, with energy $E_{\mathrm {e}}$.  The $x$ and $y$ axes
are chosen perpendicular to $Oz$ in such a way that this particle has
no $y$ component of velocity. The position of the $e^{\pm}$ is then
given by coordinates $(x_{\mathrm {e}},y_{\mathrm {e}},z_{\mathrm
{e}})$ with $\overline y_{\mathrm {e}}=0$ and $\overline x_{\mathrm
{e}}$ given in
\cite{hillas}. 
In Fig.~\ref{fig:calc}, point $E$ represents the $e^{\pm}$'s  mean
position and the circle centered on $O'$ corresponds to Cherenkov light
emitted from point $E$. As explained in section 3, the telescope mirror
is allowed to rotate around $O$ at distance $D$ and the corresponding 
circle intersects the Cherenkov light circle at $T$ and $T'$  which are
the telescope locations from which the $e^{\pm}$ contributes to the
image. Consider the location $T$ and denote by $\Phi$ the angle
$(Ox,OT)$; $OT$ gives the direction of the image axis in the focal
plane and the angular coordinates in this plane (in the longitudinal
and transverse direction respectively) are given by: 
\[
\xi_{\mathrm {l}} = \frac{D - x_{\mathrm {e}} \cos \Phi - 
y_{\mathrm {e}} \sin \Phi}{z_{\mathrm {e}}-z_{\mathrm {T}}} \;\;\;\;\;\;
\xi_{\mathrm {t}} = \frac{ x_{\mathrm {e}} \sin \Phi - y_{\mathrm {e}} 
\cos \Phi}{z_{\mathrm {e}}-z_{\mathrm {T}}}
\]
where $z_{\mathrm {T}}$ is the $z$-coordinate of the telescope. In
order to include the  contribution of position fluctuations, a
semi-empirical procedure has been used.  If the charged particle
position ($x_{\mathrm {e}}$, $y_{\mathrm {e}}$) is allowed to have a
spread around $E$ of $\sigma_x$ and $\sigma_y$ (as given in
\cite{hillas}), the Cherenkov light circle does not always intersect
the circle centered on $O$ with radius $D$; this results in an apparent
reduction of $\sigma_x$ and $\sigma_y$. Therefore, the image transverse
spread $\sigma_{\mathrm {t}}$ due to the preceding $e^{\pm}$'s is
evaluated as follows:
\[
\sigma_{\mathrm {t}}^2 = \alpha^2 \frac{(\overline x_{\mathrm {e}}^2+
\sigma_x^2) \sin^2 \Phi + 
\sigma_y^2 \cos^2 \Phi}{(z_{\mathrm {e}}-z_{\mathrm {T}})^2}
\]
where $\alpha$ has been adjusted to the value $\simeq 0.5$ in order to
agree with the results of Monte-Carlo calculations.  The form of this
semi-empirical formula is not critical to the method, since the pixel
size is much greater than $\sigma_{\mathrm {t}}$.

In the calculation of the longitudinal profile, $\xi_{\mathrm {l}}$ is
calculated for $x_{\mathrm {e}} = \overline x_{\mathrm {e}}$ and
$y_{\mathrm {e}} = 0$. The linear density of Cherenkov light along the
image axis is given by summing over the $e^{\pm}$'s, i.e. over
$\theta_{\mathrm {e}}$, $E_{\mathrm {e}}$, and $z_{\mathrm {e}}$ at
fixed $\xi_{\mathrm {l}}(\theta_{\mathrm {e}},E_{\mathrm
{e}},z_{\mathrm {e}})$:
\[
\frac{dN}{d\xi_{\mathrm {l}}} = 
\int\!\!\int\!\!\!\!\!\!\!\!\!\!\!\!\!\!\!\:
\int_{\xi_{\mathrm {l}}=\xi_{\mathrm {l}}
(\theta_{\mathrm {e}},E_{\mathrm {e}},z_{\mathrm {e}})} \, 
\frac{\partial^2 N_{\mathrm {e}}}{\partial \theta_{\mathrm {e}} 
\partial E_{\mathrm {e}}} \, 
\frac{{\mathrm{d}}L}{{\mathrm{d}}z_{\mathrm {e}}} \, 
P \, {\mathrm{d}}\theta_{\mathrm {e}} \, {\mathrm{d}}E_{\mathrm {e}} 
\, {\mathrm{d}}z_{\mathrm {e}} 
\]
where $\partial^2 N_{\mathrm {e}}/(\partial \theta_{\mathrm {e}} 
\partial E_{\mathrm {e}})$ gives the
angular and energy distrib\-utions of the $e^{\pm}$'s, 
${\mathrm{d}}L/{\mathrm{d}}z_{\mathrm {e}}$ is the
amount of Cherenkov light emitted per unit track length and $P$ is the
fraction of this light which, on average, reaches the mirror:
\[
P = \frac{\delta\omega}{\pi} \, \frac{S}{4 \pi D R}
\]
where the first ratio is the fraction of Cherenkov light impinging on
the mirror at $T$ (the angle $\delta\omega$ is defined in 
Fig.~\ref{fig:calc}) and the second the probability to find the mirror
(with area $S$ and radius $R$) around $T$. Similarly, the square of the
overall transverse spread of the image  $\Sigma_{\mathrm {t}}$ for a
given value of $\xi_{\mathrm {l}}$ is obtained by averaging 
$\sigma_{\mathrm {t}}^2(\theta_{\mathrm {e}},E_{\mathrm {e}},z_{\mathrm
{e}})$ over $\theta_{\mathrm {e}}$, $E_{\mathrm {e}}$, and $z_{\mathrm
{e}}$ at fixed $\xi_{\mathrm {l}}(\theta_{\mathrm {e}},E_{\mathrm
{e}},z_{\mathrm {e}})$. The light calculated in a given interval of
$\xi_{\mathrm {l}}$ is considered to be distributed in the transverse
direction as indicated by the following formula, scaled to the
calculated spread $\Sigma_{\mathrm {t}}$:
\[
F(\xi_{\mathrm {t}}) = \frac{1}{\sqrt{2} \Sigma_{\mathrm {t}}} 
\exp(-\frac{\sqrt{2}\mid \xi_{\mathrm {t}} \mid}{\Sigma_{\mathrm {t}}})
\;\;\;\;.
\]

\begin{ack}
The authors are grateful to the French and Czech ministries of Foreign
Affairs for providing grants for physicists' travel and accommodation
expenses.
\end{ack}

\end{document}